\documentstyle[emulateapj,psfig,apjfonts]{article}
\input psfig.sty
\submitted{Received 2002 Jul 28; Accepted: 2002 Aug 22}

\def\gs{\mathrel{\raise0.35ex\hbox{$\scriptstyle >$}\kern-0.6em
\lower0.40ex\hbox{{$\scriptstyle \sim$}}}}
\def\ls{\mathrel{\raise0.35ex\hbox{$\scriptstyle <$}\kern-0.6em
\lower0.40ex\hbox{{$\scriptstyle \sim$}}}}

\makeatother

\lefthead{Ledlow et al.}
\righthead{SCUBA galaxies behind A\,851}

\begin{document}

\title{GMOS Spectroscopy of SCUBA galaxies behind A\,851}

\author{
M.\,J.\ Ledlow,\altaffilmark{1} Ian Smail,\altaffilmark{2} 
F.\,N.\ Owen,\altaffilmark{3} W.\,C.\ Keel,\altaffilmark{4} 
R.\,J.\ Ivison\altaffilmark{5} \& G.\,E.\ Morrison,\altaffilmark{6}
}
\altaffiltext{1}{Gemini Observatory, Southern Operations Center, AURA,
Casilla 603, La Serena, Chile}
\altaffiltext{2}{Institute for Computational Cosmology, Department of Physics,
University of Durham, South Road, Durham DH1 3LE}
\altaffiltext{3}{National Radio Astronomy Observatory, P.\ O.\ Box O,
Socorro, NM 87801 USA}
\altaffiltext{4}{Dept. of Physics \& Astronomy, University of Alabama,
Tuscaloosa, AL 35487 USA}
\altaffiltext{5}{Astronomy Technology Centre, Royal Observatory, Blackford Hill, Edinburgh EH9 3HJ}
\altaffiltext{6}{California Institute of Technology, IPAC, MS\,100-22,
Pasadena, CA 91125 USA}

\setcounter{footnote}{6}

\begin{abstract}
  We have identified counterparts to two submillimeter (submm)
  sources, SMM\,J09429+4659 and SMM\,J09431+4700, seen through the
  core of the $z=0.41$ cluster A\,851. We employ deep 1.4-GHz
  observations and the far-infrared/radio correlation to refine the
  submm positions and then optical and near-infrared imaging to locate
  their counterparts.  We identify an extremely red counterpart to
  SMM\,J09429+4659, while GMOS spectroscopy with Gemini-North shows
  that the $R=23.8$ radio source identified with SMM\,J09431+4700 is a
  hyperluminous infrared galaxy (L$_{FIR}\sim 1.5\times
  10^{13}$L$_\odot$) at $z=3.35$, the highest spectroscopic redshift
  so far for a galaxy discovered in the submm. The emission line
  properties of this galaxy are characteristic of a narrow-line
  Seyfert-1, although the lack of detected X-ray emission in a deep
  {\it XMM-Newton} observation suggests that the bulk of the
  luminosity of this galaxy is derived from massive star formation.
  We suggest that active nuclei, and the outflows they engender, may
  be an important part of the evolution of the brightest submm
  galaxies at high redshifts.
\end{abstract}


\section{Introduction}

Sensitive surveys in the submm and millimeter wavebands have
identified a population of distant dusty, active galaxies which may
represent the formation phase of massive spheroidal galaxies (Smail,
Ivison \& Blain 1997; Hughes et al.\ 1998; Bertoldi et al.\ 2000;
Scott et al.\ 2002; Webb et al.\ 2002).  Irrespective of the precise
mechanism responsible for the prodigious luminosity of these galaxies,
either star formation or dust-reprocessed radiation from an AGN,
several have been confirmed as high redshift massive, gas-rich
galaxies (Frayer et al.\ 1998, 1999).  One of the most pressing issues
for study is to identify the most distant examples.  These provide a
critical test of theoretical models of galaxy formation and evolution,
which already struggle to produce sufficiently large gas masses in
galaxies at $z\sim 2$ (Granato et al.\ 2002).  Identifying similarly
luminous, gas- and dust-rich mergers at even higher redshifts, $z>3$,
will provide even stronger constraints.

In this letter we discuss the identification and spectroscopic
follow-up of two recently discovered submm galaxies in the field of
the $z=0.41$ cluster A\,851.  Exploiting very deep radio and
optical/near-infrared images we identified counterparts to both submm
sources and subsequently targetted them in spectroscopic observations
with the GMOS spectrograph on Gemini-North.  One of these galaxies has
the highest spectroscopic redshift for a SCUBA galaxy to date, at
$z=3.35$.  We adopt a cosmology with $\Omega_{\rm m}=0.3$,
$\Omega_\Lambda=0.7$ and H$_{\rm o}=70$\,km\,s$^{-1}$\,Mpc$^{-1}$.

%
%
\begin{figure*}[tbh]
\centerline{\psfig{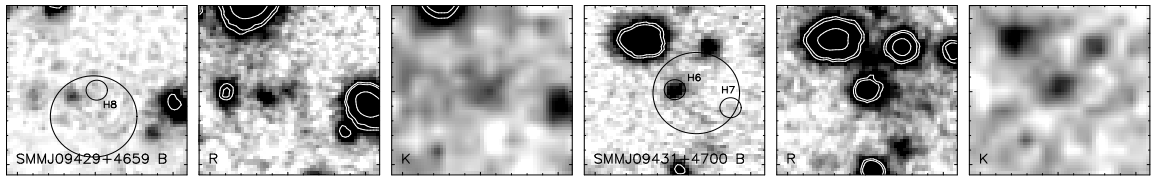}}
\caption{$BRK$ views of the fields containing the
  submm sources SMM\,J09429+4659 and SMM\,J09431+4700.  The large
  circles show the nominal $6''$-diameter error circle for the SCUBA
  sources, while the smaller circles show the positions of the radio
  counterparts.  Note the strong contrast between the
  optical/near-infrared colors of H6 and the Extremely Red Object, H8.
  Each panel is $12''\times 12''$ with North top
  and East left.}
\end{figure*}
\section{Observations and Reduction}

The two SCUBA galaxies discussed here were identified by Cowie, Barger
\& Kneib (2002) in a deep 850$\mu$m SCUBA map of A\,851
($\sigma=0.8$\,mJy).  These observations also detected
SMM\,J09429+4658, previously catalogued and studied by Smail et al.\
(1999, 2002a). The two new sources are SMM\,J09431+4700 (09\,43\,03.96
+47\,00\,16.0, all coordinates are J2000), which has an 850$\mu$m flux
density of 10.5mJy, and SMM\,J09429+4659 (09\,42\,53.49 +46\,59\,52.0)
at 4.9-mJy.  Cowie et al.\ (2002) use a detailed lens model for the
cluster to estimate that these sources are likely to be amplified by a
factor of 1.2 and 1.3$\times$ respectively.

The 1.4-GHz VLA map used in our analysis reaches a 1-$\sigma$ noise
level of 3.5\,$\mu$Jy\,beam$^{-1}$ in the field center.  Full details
of the radio observations, their reduction and cataloging are given in
Owen et al.\ (2002) and Smail et al.\ (2002b).  The radio source
surface density down to a 5-$\sigma$ limit of 17.5$\mu$Jy is
4\,arcmin$^{-2}$ (Smail et al.\ 2002b).  Adopting a nominal error
radius of 3$''$ for the SCUBA sources (Smail et al.\ 2002a), this
translates into only a 3\% chance of an unrelated radio source falling
within the SCUBA error circle.  Searching within this radius around the
position of SMM\,J09429+4659 we identify a bright, unresolved
($<$0.15$''$) 970\,$\mu$Jy radio source: H8 [09\,42\,53.42
+46\,59\,54.5], 2.5$''$ from the submm position.  For SMM\,J09431+4700
we identify two radio counterparts; one has a 1.4-GHz flux density of
72\,$\mu$Jy (H6: 09\,43\,04.08 +47\,00\,16.2) and is 1.2$''$ from the
submm position, the other has 55\,$\mu$Jy (H7: 09\,43\,03.70
+47\,00\,15.1) and is 2.8$''$ away.  The low surface density of radio
sources suggests that H6 and H7 are both related to the submm emission.
Both of these sources are slightly resolved in our VLA map, with sizes
of 0.7--0.9$''$.

We next exploit deep $BV\! RIzJHK$ imaging of this field to identify
the radio source counterparts of the submm galaxies in the optical and
near-infrared.  These images reach typical 3-$\sigma$ depths of
$m_{AB}\sim 27$ in the optical and $m_{AB}\sim 24$ in the
near-infrared and their reduction and calibration are described by
Kodama et al.\ (2001) and Smail et al.\ (2002b).  We astrometrically
align the near-infrared images to the radio frame with a precision of
0.32$''$ rms using the positions of 90 sources detected in the two
wavebands.  This allows us to pinpoint near-infrared counterparts to
the many faint radio sources in our VLA map.

All three of the radio counterparts to the SCUBA galaxies are
identified in our optical or near-infrared data: H6 has $K=20.2$ and
$(R-K)=3.59\pm 0.16$, H7 is undetected in $K$ but is coincident with a
$R=25.9$, $(R-K)\leq 4.6$ galaxy, and H8 has $K=19.7$ and is extremely
red $(R-K)=5.5\pm 0.3$.  We show the optical/near-infrared images of
these galaxies in Fig.~1.  Unfortunately, neither galaxy lies in the
sparse {\it HST} mosaic of this field discussed by Dressler et al.\ (in
prep) and only H6 is detected at sufficient signal-to-noise in our
good-seeing optical images to reliably measure its intrinsic FWHM:
$0.4\pm 0.1''$.

The spectroscopy of these galaxies was attempted with the GMOS
spectrograph (Davies et al.\ 1997) on Gemini-North.  The target of
this survey was faint radio galaxies, which are likely to lie at
$z\sim 1$--4 (Owen et al.\ 2002), and so we aimed for the widest
possible wavelength coverage and the highest resolution to maximise
the chance of detecting weak emission and absorption features in these
optically-faint galaxies.  To accomplish this we observed each mask
twice, once with the B600 grating (centered at 4730\AA) and then with
the R400 grating (centered at 7920\AA) to provide continuous spectral
coverage from 3500\AA\ to 1$\mu$m. The detectors were read out in 3-amp
mode with $2\times 2$ binning (along both the spatial and dispersion
axes). The resulting pixel scale is 0.145$''$ pixel$^{-1}$ with
dispersions of 0.91\AA\ pixel$^{-1}$ (B600) and 1.37\AA\ pixel$^{-1}$
(R400). Slitlets of 1$'' \times 5''$ were placed on each target,
giving a resolution of 3--4\AA.

A multislit mask was designed for the center field in A\,851 which
includes both submm sources.  A total of 2.5 hrs integration was
obtained with each of the B600 and R400 gratings on the nights of March
14 and 15, 2002 in 0.7--0.8$''$ seeing.  These observations were
reduced and calibrated in a standard manner using {\sc iraf} scripts.
We used a CuAr lamp spectrum for wavelength calibration, and removed
any remaining offsets using bright sky lines. The spectra were flux
calibrated with observations of Feige 34 (B600) and HZ\,44 (R400)
observed through a 1$''$ long-slit.  Both H6 and H8 were targeted in
the same mask.  H7's proximity to H6 precluded placing slitlets on both
sources.  ~From the observation of the source H8 ($R=25.2$), we are
unable to identify either any features or detect the continuum.
However, we do identify a series of strong emission lines in the
spectrum of H6 shown in Fig.~2.

Finally, we have retrieved the {\it XMM/Newton} X-ray image of this
field from the public archive\footnote{This work is based on
observations obtained with the {\it XMM-Newton}, an ESA science mission
with instruments and contributions directly funded by ESA member states
and the USA (NASA).} to place limits on the hard X-ray emission from
the submm sources.  This image provides 49.4-ks of useful integration
from the EPIC MOS1 and MOS2 cameras and allows us to place 3-$\sigma$
limits of $0.9\times 10^{-15}$\,ergs\,s$^{-1}$ and $1.3\times
10^{-15}$\,ergs\,s$^{-1}$ on the unabsorbed 2--10\,keV fluxes of
SMM\,J09429+4659 and SMM\,J09431+4700.  Adopting a photon index of 1.7,
these limits translate into 3-$\sigma$ lower bounds on the submm-X-ray
spectral indices (Fabian et al.\ 2000) of $>1.23$ and $>1.25$
respectively.

%
%
\begin{figure*}
\centerline{\psfig{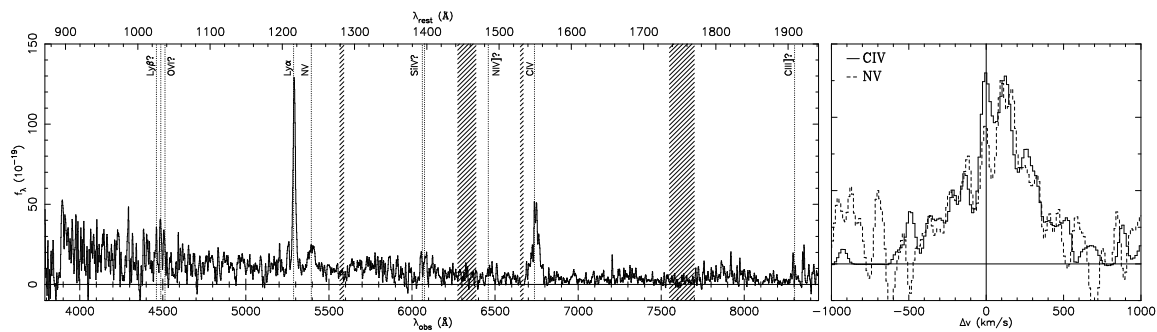}}
\caption[]{The left-hand panel shows the GMOS spectrum of
SMM\,J09431+4700 with the strongest features marked.  The
cross-hatched regions indicate areas effected by detector features or
atmospheric absorption or emission.  The spectral properties of this
galaxy are similar to those seen in narrow line Seyfert-1's. There are
also hints of blue-shifted absorption troughs on the stronger lines,
although this is only tentative given the modest signal-to-noise of
our observations.  The right-hand panel compares the morphologies of
the N{\sc v} and C{\sc iv} emission lines plotted on a
restframe velocity scale relative to the redshift of the narrow
Ly$\alpha$ line, the line centres are shown by the
vertical solid line (the horizontal solid line indicates the continuum
level). The matches between the different components probably reflects
structure in the emission region.  A 30\,km\,s$^{-1}$ shift
to the blue has been applied to the N{\sc v} line, this is within the
relative calibration error of the red and blue spectra.  To aid this
comparison, both spectra have been smoothed to the instrumental
resolution with a 3\AA\ FWHM Gaussian.}
\end{figure*}

\section{Analysis and Results}

We identify several strong emission lines in the combined GMOS spectrum
of H6 shown in Fig.~2, including Ly$\alpha$, N{\sc v} and C{\sc iv}.
Based on the wavelength of the narrow Ly$\alpha$ line we estimate a
redshift of $z=3.349$, consistent with the observed wavelengths of the
other broader emission features.  This is currently the highest
spectroscopic redshift for the counterpart of a SCUBA galaxy --
breaking the $z=3$ barrier for this population for the first time.
While this is only a modest increase in lookback time, compared to the
previously most-distant source at $z=2.80$, it corresponds to a
decrease of a factor of two in the predicted abundance of the massive
halos, $> 10^{12}$\,M$_\odot$, believed to host these luminous galaxies
(Jenkins et al.\ 2001).  The exponential decline in the space density
of the most massive halos at higher redshifts underlines the strong
constraints available from identifying the highest redshift submm
galaxies.

The 850$\mu$m flux of SMM\,J09431+4700 of 8.8\,mJy (corrected for
lensing) translates into L$_{FIR}\sim 1.5 \times 10^{13}$\,L$_\odot$,
assuming T$_d=38$\,K.\footnote{If {\it both} H6 and H7 contribute to
the submm emission then from their joint radio flux we would estimate a
somewhat higher dust temperature, T$_{~d}=49$\,K, and a far-infrared
luminosity of L$_{FIR}=3.6\times 10^{13}$\,L$_\odot$.}  This makes this
galaxy a Hyperluminous Infrared system (HyLIRG, Rowan-Robinson
2000). If purely powered by massive star formation, its immense
luminosity would require a star formation rate of $\sim
10^4$\,M$_\odot$\,yr$^{-1}$!

However, the spectrum of H6 shows the signatures of a weak AGN.  The
strongest lines include a narrow and symmetrical Ly$\alpha$ line with a
flux of $1.9\times 10^{-16}$\,ergs\,s$^{-1}$\,cm$^{-2}$ and a FWZI of
55\AA.  There are also broader emission lines coincident with the
expected wavelength of redshifted N{\sc v}$\lambda$\,1239.7, with an
observed FWZI of $\sim 155$\AA, and an integrated flux of $\sim
0.7\times 10^{-16}$\,ergs\,s$^{-1}$\,cm$^{-2}$, and C{\sc
iv}$\lambda$\,1549, with a FWZI of $\sim 173$\AA\ and a flux of
$2.1\times 10^{-16}$\,ergs\,s$^{-1}$\,cm$^{-2}$.  We also identify a
number of weaker emission lines and mark these in Fig.~2.  The velocity
widths of these lines cover a wide range, from a restframe FWHM of
210\,km\,s$^{-1}$ for Ly$\alpha$ upto $\sim$ 550\,km\,s$^{-1}$ for the
higher excitation lines.  In addition the morphologies of the N{\sc v}
and C{\sc iv} lines appear very similar (Fig.~2).  The spectral
properties, line widths and line ratios of this galaxy are very similar
to those seen for narrow line Seyfert-1's (NLSy1, Osterbrock \& Pogge
1985; Crenshaw et al.\ 1991), in particular Mrk\,24. This
classification is supported by the ratio He{\sc ii}$\lambda$1640/C{\sc
iv}=0.05, indicative of a NLSy1 or NLQSO (Heckman et al.\ 1995).  These
galaxies have narrow forbidden and permitted emission lines, FWHM\,$\ls
400$--700\,km\,s$^{-1}$, which are thought to result from an increase
in the size and density of the broad line region compared to a normal
Seyfert-1 (Laor et al.\ 1997).

Based on our deep imaging we measure 3$''$-diameter photometry of H6
giving: $B=25.82\pm 0.07$, $V=24.11\pm 0.03$, $R=24.00\pm 0.03$,
$I=23.62\pm 0.07$, $z>24.1 [3\sigma]$, $J>22.7 [3\sigma]$, $H=21.41\pm
0.25$, $K=20.41\pm 0.14$.  The $(B-V)$ color is relatively red,
$(B-V)=1.7$, suggesting the presence of either strong absorption in the
$B$-band or strong emission in the $V$-band.  We note that at $z=3.35$
the Lyman-limit falls just blueward of the $B$-band (indeed emission is
seen down to 890\AA\ in the restframe, rising as $\lambda^{-2.5}$),
however, several strong lines fall in the $V$-band (Fig.~2).
Integrating all the flux in the $V$-band part of the GMOS spectrum we
measure a slit magnitude of $V\sim 24.4$, in reasonable agreement with
the 3$''$-diameter aperture magnitude and from this we estimate that
50\% of the $V$-band light is contributed by Ly$\alpha$ and N{\sc v}, a
similar calculation for C{\sc iv} suggests that it contributes 35\% of
the $R$-band flux.  Using the apparent $K$-band magnitude of this
galaxy we estimate an absolute $V$-band magnitude of $M_V\sim -23.0$ at
$z=3.35$, although a significant fraction of this light may be
contributed by emission lines, including H$\beta$ and [O{\sc
iii}]\,$\lambda$4959,5007.

We also investigate the spatial extent of the emission lines along our
slit -- focusing on Ly$\alpha$ and N{\sc v} as these lines lie close in
wavelength. We find that Ly$\alpha$ is spatially extended compared to
N{\sc v} with a FWHM of 0.95$''$ versus 0.74$''$. As the estimated
image quality in the $r$-band was 0.7$''$ at the beginning of the
observations, the N{\sc v} width is consistent with being spatially
unresolved. For Ly$\alpha$ we measure an intrinsic FWHM of $\sim 0.6''$
-- indicating extended Ly$\alpha$ emission on a scale of 8\,kpc.

Turning to the radio properties of this galaxy, the radio luminosity
of H6 is $5\times 10^{24}$\,W\,Hz$^{-1}$ for an $\alpha=-1$ spectral
slope or $2.5\times 10^{24}$\,W\,Hz$^{-1}$ for $\alpha=-0.7$
(including a 20\% correction for lensing).  These luminosities
correspond locally to those expected for a FR I radio galaxy, although
it is equally consistent with strong radio emission from a massive
starburst at the level detected in the submm waveband.  The
radio-submm spectral index, $\alpha^{850}_{1.4}$, is $0.91\pm 0.04$
based on the radio emission from H6 (if H7 contributes as well then
$\alpha^{850}_{1.4}\sim 0.80$).  These correspond to a redshift of
$z=2.7_{-0.9}^{+1.7}$ using the models from Carilli \& Yun (2000), in
reasonable agreement with our spectroscopic measurement.\footnote{For
  the ERO/mJy-radio counterpart to SMM\,J09429+4659, H8, we estimate
  $\alpha^{850}_{1.4}=0.30\pm 0.07$, corresponding to only $z=0.4\pm
  0.3$.  It is clear that the radio emission must include a
  significant contribution from an AGN and the estimated redshift is
  only a lower limit -- assuming the galaxy follows the $K$--$z$
  relationship for powerful radio galaxies it most likely lies at
  $z\gs 2$, Jarvis et al.\ (2001).}

The {\it XMM-Newton} observations suggest that unless highly obscured,
the intrinsic luminosity of the AGN in H6 must be modest,
$L_X($2--10\,keV$)\ls 10^{42}$\,ergs\,s$^{-1}$.  However, the limit on
the submm-X-ray spectral index of H6, $>1.25$, is also consistent with
a heavily obscured AGN such as NGC\,6240 (Keel 1990), with N(H{\sc
i})\,$\sim 10^{24}$, at $z=3.35$ (Fabian et al.\ 2000).

\section{Discussion and Conclusions}

The spectroscopic identification of a second HyLIRG in
the SCUBA population provides the opportunity for a detailed
comparison of the properties of these two galaxies: SMM\,J02399$-$0136
and SMM\,J09431+4700.  Detailed study of SMM\,J02399$-$0136 has shown
that it is a massive, gas-rich system (Ivison et al.\ 1998; Frayer et
al.\ 1998), identified with a pair of galaxies: L1 and L2, at
$z=2.80$, and has a far-infrared luminosity of L$_{FIR}\sim 1\times
10^{13}$\,L$_\odot$.  L1 hosts a partially obscured AGN, which has
recently been classified as a Broad Absorption Line (BAL) QSO (Vernet
\& Cimatti 2001).  The second component L2 is separated from L1 by
about 10\,kpc in projection (compared to 50\,kpc for H6--H7), and may
be tidal debris rather than an independent galaxy.

Apart from their apparent binary morphologies, the most striking
similarity between these two HyLIRGs is that both systems host AGN.
As highlighted by Ivison et al.\ (2000): 80\% of the SCUBA galaxies
with known redshifts show some signs of an AGN.  The close
relationship of AGN and QSO activity to the growth of supermassive
black holes (SMBH) and the apparent ubiquity of SMBH in local, massive
spheroids suggests that this is a natural consequence if SCUBA
galaxies are the progenitors of the most massive galaxies in the local
Universe (e.g.\ Sanders et al.\ 1988; Silk \& Rees 1998; Granato et
al.\ 2001; Archibald et al.\ 2001).  One measure of the importance of
the AGN in these systems is to quantify its contribution to their
total emission. X-ray observations of the two HyLIRG submm galaxies
suggests that in neither does the AGN dominate the bolometric emission
(Bautz et al.\ 2000), which instead comes from an intense starburst
which also produces the substantial masses of dust in these galaxies.

Although they do not dominate the energetics, the AGN may still have a
profound effect on the evolution of these galaxies: the AGN in
SMM\,J02399$-$0136 is driving a substantial wind which may in time
sweep the central regions of the galaxy clear of gas and dust.  Can we
find any signs of similar AGN-induced feedback in SMM\,J09431+4700?
There are several hints: the contrast between the narrow,
spatially-extended Ly$\alpha$ emission and the broader, but
spatially-unresolved high excitation lines, may indicate that the
former arises from emission in an outflow.  This situation is very
similar to that seen in some high-redshift radio galaxies, such as
53W002 at $z=2.4$ (Windhorst, Keel \& Pascarelle 1998).  Moreover, the
structured broad emission lines seen in H6 are reminiscent of the
structures seen in broad emission lines of some radio galaxies and
radio-loud quasars (Eracleous \& Halpern 1994), although those show
much larger velocity ranges.  These structured emission lines are
interpreted as resulting from scattering of radiation from the AGN by
outflowing conical winds (Corbett et al.\ 1998) and the same mechanism
may be operating in SMM\,J09431+4700/H6.  The final connection is from
the spectral classification of this galaxy as a NLSy1, where analysis
of examples at $z\sim 0$ have led to the identification of strong
nuclear winds and a suggested link to BALQSOs (Lawrence et al.\ 1997;
Leighly et al.\ 1997; Laor et al.\ 1997).

If both systems do show signatures of massive outflows -- this
suggests that winds powered by the AGN (as well as starbursts) must be
central to our understanding the growth of the spheroidal components
in these massive, young galaxies (Granato et al.\ 2002).  The feedback
on the system from energy injected by the AGN provides one possible
mechanism for creating the observed SMBH to bulge mass correlation
seen in local galaxies (Maggorian et al.\ 1998).  The future evolution
of these submm sources will be determined by the ability of the AGN to
clear the bulk of the gas and dust from the nuclear regions -- if they
can, then they may evolve into the population of less-obscured QSOs.

With regard to the wider environment of H6, we note a striking
coincidence: H6 is just 400\,km\,s$^{-1}$ and less than
1\,Mpc ($64''$) from an optically-selected galaxy, DG\,433
($z=3.3435$), (Trager et al.\ 1997), suggesting that the two galaxies
inhabit a single structure.  DG\,433 has a UV spectrum dominated by
absorption lines and an estimated star formation rate of $\ls
10^2$\,M$_\odot$\,yr$^{-1}$, indicating it is a much less active system than
the hyperluminous galaxy H6.  The relationship between the highly
obscured and very active submm-luminous galaxies and the less obscured
populations which cluster in the same environments will be one of the
most important questions to address in the next few years.

\acknowledgments

Based on observations obtained at the Gemini Observatory, which is
operated by the Association of Universities for Research in Astronomy,
Inc., under a cooperative agreement with the NSF on behalf of the
Gemini partnership: the National Science Foundation (United States),
the Particle Physics and Astronomy Research Council (United Kingdom),
the National Research Council (Canada), CONICYT (Chile), the
Australian Research Council (Australia), CNPq (Brazil) and CONICET
(Argentina).  We thank Taddy Kodama for the use of his exquisite
Subaru imaging and an anonymous referee for helpful comments.  We also
thank Andrew Blain, Chris Carilli, Scott Chapman, Len Cowie, Chris
Done, Alastair Edge, Inger Jorgensen, Jean-Paul Kneib, Rowena Malbon,
Bryan Miller, Matt Page, Graham Smith for useful discussions and help.
IRS acknowledges support from Royal Society and Philip Leverhulme
Prize Fellowships.

\end{document}